\documentclass[aps,twocolumn,superscriptaddress]{revtex4-1}
\usepackage[colorlinks=true,bookmarks=true,citecolor=magenta,urlcolor=magenta,linkcolor=magenta,breaklinks]{hyperref} 
\usepackage{breakurl}
\usepackage{sidecap}
\usepackage{amssymb}
\usepackage{hhline}
\usepackage{urwchancal}
\usepackage{xcolor}
\sidecaptionvpos{figure}{t}
\usepackage{amsmath}
\usepackage{graphicx}
\usepackage{esint}
\usepackage{epstopdf}
\usepackage{rotating}
\graphicspath{{pict/}{./}}
\usepackage{bm}%
\usepackage{microtype,bm,bbm,graphicx,booktabs,times}

\newcounter{Fig}

\begin{document}


\title{Momentum-Space Scattering Extremizations}
\author{Chunchao Wen}
\affiliation{College for Advanced Interdisciplinary Studies, National University of Defense Technology, Changsha 410073, P. R. China.}
\affiliation{Nanhu Laser Laboratory and Hunan Provincial Key Laboratory of Novel Nano-Optoelectronic Information Materials and
Devices, National University of Defense Technology, Changsha 410073, P. R. China.}
\author{Jianfa Zhang}
\email{jfzhang85@nudt.edu.cn}
\affiliation{College for Advanced Interdisciplinary Studies, National University of Defense Technology, Changsha 410073, P. R. China.}
\affiliation{Nanhu Laser Laboratory and Hunan Provincial Key Laboratory of Novel Nano-Optoelectronic Information Materials and
Devices, National University of Defense Technology, Changsha 410073, P. R. China.}
\author{Shiqiao Qin}
\affiliation{College for Advanced Interdisciplinary Studies, National University of Defense Technology, Changsha 410073, P. R. China.}
\affiliation{Nanhu Laser Laboratory and Hunan Provincial Key Laboratory of Novel Nano-Optoelectronic Information Materials and
Devices, National University of Defense Technology, Changsha 410073, P. R. China.}
\author{Zhihong Zhu}
\affiliation{College for Advanced Interdisciplinary Studies, National University of Defense Technology, Changsha 410073, P. R. China.}
\affiliation{Nanhu Laser Laboratory and Hunan Provincial Key Laboratory of Novel Nano-Optoelectronic Information Materials and
Devices, National University of Defense Technology, Changsha 410073, P. R. China.}
\author{Wei Liu}
\email{wei.liu.pku@gmail.com}
\affiliation{College for Advanced Interdisciplinary Studies, National University of Defense Technology, Changsha 410073, P. R. China.}
\affiliation{Nanhu Laser Laboratory and Hunan Provincial Key Laboratory of Novel Nano-Optoelectronic Information Materials and
Devices, National University of Defense Technology, Changsha 410073, P. R. China.}

\begin{abstract}
Studies into scatterings of photonic structures have been so far overwhelmingly focused on their dependencies on the spatial and spectral morphologies of the incident waves. In contrast, the evolution of scattering properties through another parameter space of incident directions (momentum space) has attracted comparably little attention, though of profound importance for various scattering-related applications. Here we investigate, from the perspective of quasi-normal modes (QNMs), the momentum-space scattering  extremizations with respect to varying incident directions of plane waves. It is revealed that for effective single-QNM excitations, scatterings are maximized exactly along those directions where the QNM radiation reaches its maximum, with matched incident and radiation polarizations. For an arbitrary direction, when the incident polarization is tuned to be orthogonal to that of the mode radiation, the QNM cannot be excited and thus the scatterer becomes invisible with null scatterings. The principles we have revealed are protected by fundamental laws of reciprocity and energy conservation (optical theorem), which can  be further expanded and applied for other branches of wave physics.
\end{abstract}

\maketitle

\section{Introduction}
Throughout all disciplines of photonics that involve light-matter interactions, scattering manipulations (enhancement or suppression \textit{e.g.} cloaking) constitute one of the central themes~\cite{Bohren1983_book,YARIV_2006__Photonics,LIU_2005__Photonic}. To control the interactions, either the incident source or the photonic structure can be engineered to satisfy different demands of various applications. For source engineering, previous studies are extensively focused on frequency tuning  and/or spatial phase and polarization morphology structuring ~\cite{ZUMOFEN_Phys.Rev.Lett._perfect_2008-1,GOUESBET_generalized_2011,SHEN_2019_LightSciAppl_Optical,WAN_2022_Natl.Sci.Rev._Photonic,FORBES_Nat.Photonics_Structured}. Though it is apparent that light-matter interactions are largely dependent on incident directions, systematic and thorough examinations of momentum-space scattering extremizations (maximized or minimized) have not yet been conducted. All possible incident directions constitute a closed momentum sphere, and according to the extreme value theorem~\cite{RUDIN_1976__Principles} there must be at least one direction along which the scattering can reach its maximum or minimum.

Here we study the momentum-space scattering extremizations with respect to varying directions of incident plane waves, from the perspective of QNMs supported by the scatterers. When only one QNM is effectively excited,  it is discovered that the maximum mode radiation directions correspond exactly to the momentum-space points where the scattering is maximized, with matched incident and radiation polarizations. Along an arbitrary direction, when the incident and mode radiation polarizations are tuned to be orthogonal, all scatterings are fully eliminated, rendering the scatterer invisible. Our revelations connecting QNM radiations and momentum-space scattering evolutions are secured by the fundamental laws of electromagnetic reciprocity and optical theorem. As a result, the framework we have established can be naturally extended to apply to other branches of wave physics, incubating applications in fields such as acoustics, water waves and microscopic matter waves.

\begin{figure}[tp]
\centerline{\includegraphics[width=8.5cm]{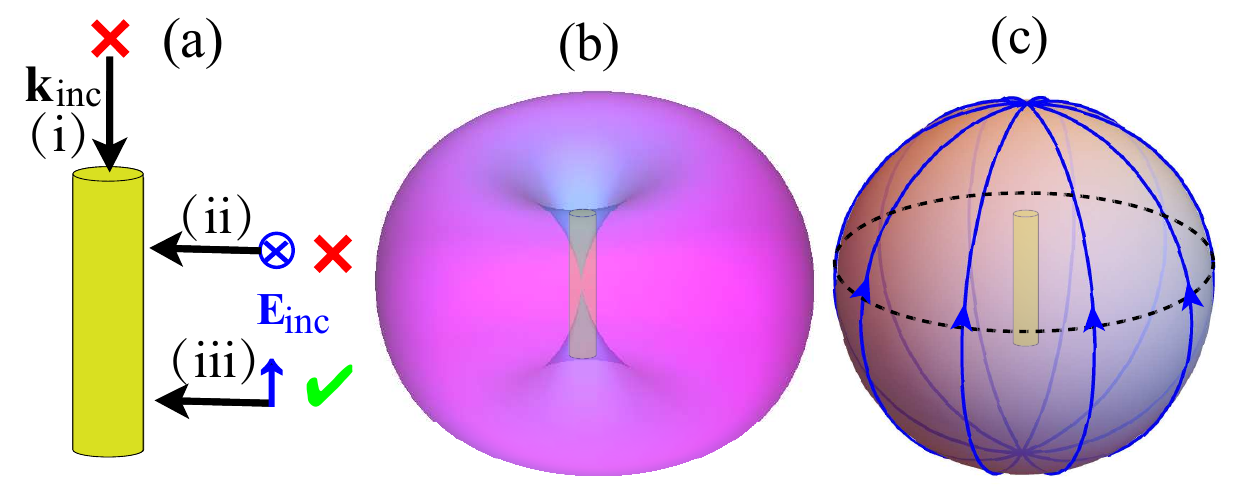}} \caption {\small (a) A metallic cylinder that supports a linear electric dipolar QNM, the angular radiation intensity and polarization distributions of which are shown in (b) and (c), respectively. Three excitation scenarios (i)-(iii) are also specified in (a), where black and blue arrows denote the propagation and electric field directions of the incident plane waves, respectively. For (i) \& (ii) the QNM cannot be  excited, while for (iii) the excitation efficiency reaches its maximum.}\label{fig1}
\end{figure} 

\section{Theoretical Model Incorporating Reciprocity and Optical Theorem}
\label{section2}

Throughout this work, we confine our studies to  the single-QNM regime: the spectral regions where only one QNM can be effectively excited, and other QNMs are either spectrally apart or simply cannot be excited. The incident sources are plane waves (electric field vector $\mathbf{E}_{\rm{inc}}$, wavevector $\mathbf{k}_{\rm{inc}}$, angular frequency $\omega$ and vacuum wavelength $\lambda$) and the QNM excited is characterized by a complex eigenfrequency $\tilde{\omega}$~\cite{LALANNE__LaserPhotonicsRev._Light}. In the single-QNM regime, all scattering properties are decided by the excitation coefficient $\alpha(\mathbf{E}_{\rm{inc}},\omega,\mathbf{k}_{\rm{inc}})$ of the QNM~\cite{CHEN_Phys.Rev.Lett._Extremize}:
\begin{equation}
\label{efficiency-dependence}
\rm C_{\rm{sca},\rm{ext},\rm{abs}} \propto 	|\alpha(\mathbf{E}_{\rm{inc}},\omega,\mathbf{k}_{\rm{inc}})|^2,
\end{equation}
where $\rm C_{\rm{sca},\rm{ext},\rm{abs}}$ are respectively scattering, extinction and absorption cross sections; 	$|\alpha(\mathbf{E}_{\rm{inc}},\omega,\mathbf{k}_{\rm{inc}})|^2$ is the excitation efficiency, and its spectral curve (dependence on $\omega$) generally exhibits a typical Lorentzian shape, with the central position and linewidth decided by the complex $\tilde{\omega}$~\cite{LALANNE__LaserPhotonicsRev._Light,CHEN_Phys.Rev.Lett._Extremize}).

A conventional approach to calculate $\alpha$ and thus the excitation efficiency relies on cumbersome volume integrations involving detailed near-field current distributions of the QNM~\cite{LALANNE__LaserPhotonicsRev._Light}.  It is recently shown that, for incident planes waves, $\alpha$ can be directly calculated in the far field according to the electromagnetic reciprocity~\cite{CHEN_Phys.Rev.Lett._Extremize}:
\begin{equation}
\label{expansion-coefficient3}
\alpha(\mathbf{E}_{\rm{inc}},\omega,\mathbf{k}_{\rm{inc}})\propto \tilde{\mathbf{E}}_{\rm{rad}} \cdot \mathbf{E}_{\rm{inc}},
\end{equation}
where $\tilde{\mathbf{E}}_{\rm{rad}}$ is the electric field vector of the QNM radiation along $\mathbf{k}_{\rm{rad}}=-\mathbf{k}_{\rm{inc}}$. To clarify the meaning of the dot product in Eq.~(\ref{expansion-coefficient3}), we employ the normalized Jones vectors $\mathbf{J_{\rm{inc,rad}}}=(j^{~\rm{inc,rad}}_1,j^{\rm{~inc,rad}}_2)$ to characterize the polarizations of the incident wave and the far-field QNM radiation, \textit{e.g.} $\mathbf{J}=\frac{1}{\sqrt{2}}(1,\pm i)$ denoting right- and left-handed circular polarizations, respectively~\cite{YARIV_2006__Photonics}. Then Eq.~(\ref{expansion-coefficient3}) can be expressed as~\cite{DEHOOP_Appl.sci.Res._reciprocitya,CHEN_2021_Phys.Rev.B_Arbitrary}:
\begin{equation}
\label{expansion-coefficient-jones}
\alpha(\mathbf{E}_{\rm{inc}}, \omega,\mathbf{k}_{\rm{inc}})\propto \mathbf{J_{\rm{rad}}}\mathbf{J^{\dagger}_{\rm{inc}}}\tilde{\rm{E}}_{\rm{rad}}\rm{{E}}_{\rm{inc}},
\end{equation}
where $\dagger$  denotes combined operations of complex conjugate and transpose; $\tilde{\rm{E}}_{\rm{rad}}$ and $\rm{{E}}_{\rm{inc}}$ are scalar field amplitudes.
It is worth mentioning that the Jones vector characterizes the polarization only and does not contain any information about the light prorogation direction, \textit{e.g.} waves of the same polarization while propagating along opposite directions are represented by the same Jones vector.

Since all cross sections $\rm C_{\rm{sca},\rm{ext},\rm{abs}}$ are defined to be independent of the incident field strength~\cite{Bohren1983_book}, Eq.~(\ref{efficiency-dependence}) can be then explicitly simplified as:
\begin{equation}
\label{efficiency-dependence-2}
\rm C_{\rm{sca},\rm{ext},\rm{abs}} \propto |\mathbf{J_{\rm{rad}}}\mathbf{J^{\dagger}_{\rm{inc}}}|^2|\tilde{\rm{E}}_{\rm{rad}}|^2=|\mathbf{J_{\rm{rad}}}\mathbf{J^{\dagger}_{\rm{inc}}}|^2{\rm{I}}_{\rm{rad}},
\end{equation}
where ${\rm{I}}_{\rm{rad}}=|\tilde{\rm{E}}_{\rm{rad}}|^2$ is the QNM radiation intensity along $\mathbf{k}_{\rm{rad}}=-\mathbf{k}_{\rm{inc}}$. For general reciprocal scatterers, it has been proved, based on the principles of reciprocity and optical theorem, that for an arbitrary pair of opposite incident directions ($\pm\mathbf{k}_{\rm{inc}}$) the extinction cross sections are identical~\cite{SOUNAS_Opt.Lett.OL_extinction_2014,CHEN_2020_Phys.Rev.Research_Scatteringa}. In the single-QNM regime, both cross sections of scattering and absorption are also identical for $\pm\mathbf{k}_{\rm{inc}}$~\cite{CHEN_Phys.Rev.Lett._Extremize}. According to Eq.~(\ref{efficiency-dependence-2}), it requires that:
\begin{equation}
\label{opposite radiation}
\mathbf{J_{\rm{rad}}}(\mathbf{k}_{\rm{rad}})= \mathbf{J_{\rm{rad}}}(-\mathbf{k}_{\rm{rad}}), ~~   {\rm{I}}_{\rm{rad}}(\mathbf{k}_{\rm{rad}})= {\rm{I}}_{\rm{rad}}(-\mathbf{k}_{\rm{rad}}).
\end{equation}
That is to say, in the single-QNM regime, the mode radiations are of the same polarization and magnitude along two arbitrary opposite directions. A more intuitive interpretation of this conclusion is as follows: (i) Equation~(\ref{efficiency-dependence-2}) tells that for an incident plane wave along $\mathbf{k}_{\rm{inc}}$, the extinction is solely decided by the QNM radiation along the opposite direction of $\mathbf{k}_{\rm{rad}}=-\mathbf{k}_{\rm{inc}}$; (ii) Optical theorem tells that the extinction is only related to QNM radiation along $\mathbf{k}_{\rm{rad}}=\mathbf{k}_{\rm{inc}}$, as only forward radiations (parallel to the incident direction) can interfere destructively with the incident wave to extinguish part of its energy to account for all-angle scatterings and dissipative absorptions~\cite{Bohren1983_book}; (i) \& (ii) are applicable for arbitrary incident polarizations, which results in Eq.~(\ref{opposite radiation}).

Now we proceed to discuss the momentum-space scattering extremizations based on Eq.~(\ref{efficiency-dependence-2}). The maximization of the scattering requires the simultaneous maximization of both ${\rm{I}}_{\rm{rad}}$ and $|\mathbf{J_{\rm{rad}}}\mathbf{J^{\dagger}_{\rm{inc}}}|$: the former corresponds to that  $\mathbf{k}_{\rm{inc}}$ is collinear to the directions along which the QNM radiation is strongest [according to Eq.~(\ref{opposite radiation}) there are at least a pair of such directions which are opposite to each other]; the latter requires the incident and QNM radiations along $\pm \mathbf{k}_{\rm{inc}}$) are of the same polarization ($|\mathbf{J_{\rm{rad}}}\mathbf{J^{\dagger}_{\rm{inc}}}|=1$), which is natural since the time reversal operation does not change the polarization of the field~\cite{DEHOOP_Appl.sci.Res._reciprocitya,CHEN_2021_Phys.Rev.B_Arbitrary}.

We have thus so far managed to reveal the recipe to obtain the maximum scattering in the single-QNM regime: (i) Identify the QNM of complex eigenfrequency $\tilde{\omega}$;  (ii) Calculate its far-field radiations, in terms of both distributions of radiation intensity ${\rm{I}}_{\rm{rad}}$ and polarization $\mathbf{J_{\rm{rad}}}$; (iii) Pinpoint the directions (at least a pair of them) where the scattering reaches its maximum and specify the polarization; (iv) Shine light along those directions with matched polarization (with the same polarization as that of the radiation: $|\mathbf{J_{\rm{rad}}}\mathbf{J^{\dagger}_{\rm{inc}}}|=1$). For a fixed incident direction, the scattering is maximized when the incident and radiation polarizations are matched; while for a fixed incident polarization, the scattering maximization requires simultaneous consideration of  $|\mathbf{J_{\rm{rad}}}\mathbf{J^{\dagger}_{\rm{inc}}}|$ and ${\rm{I}}_{\rm{rad}}$.

Equation~(\ref{efficiency-dependence-2}) also shows that along arbitrary directions, the scattering can be always fully eliminated $\rm C_{\rm{sca},\rm{ext},\rm{abs}}=0$: along directions where the QNM radiation is not zero, the elimination is obtained when the incident and radiation polarizations are orthogonal ($\mathbf{J_{\rm{rad}}}\mathbf{J^{\dagger}_{\rm{inc}}}=0)$; along directions of vanishing  radiations (${\rm{I}}_{\rm{rad}}=0$), there are no scatterings for incident waves of arbitrary polarizations.

The principles discovered and explicitly described above are schematically exemplified in Fig.~\ref{fig1}: a metallic cylinder [Fig.~\ref{fig1}(a)] supports a linear electric dipolar QNM, the angular radiation intensity and polarization (linear polarization everywhere along the longitude, except for the poles where the radiations are singularly zero~\cite{CHEN_Phys.Rev.Lett._Singularities}) distributions of which are shown respectively in Fig.~\ref{fig1}(b) and Fig.~\ref{fig1}(c). Three excitation scenarios are also included in
Fig.~\ref{fig1}: for (i) \& (ii) the excitation efficiency is zero while for (iii) it is maximized. This can be easily interpreted from the conventional perspective of near-field interactions between  $\mathbf{E}_{\rm{inc}}$ and the oscillating electrons along the cylinder axis. Meanwhile, Eq.~(\ref{efficiency-dependence-2}) provides an alternative far-field interpretation for each scenario: (i) ${\rm{I}}_{\rm{rad}}=0$ and thus $\alpha=0$; (ii) $\mathbf{J_{\rm{rad}}}\mathbf{J^{\dagger}_{\rm{inc}}}=0$ and thus $\alpha=0$; (iii) both $|\mathbf{J_{\rm{rad}}}\mathbf{J^{\dagger}_{\rm{inc}}}|$ and $\tilde{\mathbf{E}}_{\rm{rad}}$ are maximal, leading to maximized coupling efficiency and thus also scattering properties. For this elementary example, the superiority of far-field interpretations based on Eq.~(\ref{efficiency-dependence-2}) is not obvious; while for more sophisticated structures where direct analysis of charge-field interactions becomes too complicated, the overwhelming simplicity and strength of our far-field model would become apparent.

\begin{figure*}[tp]
\centerline{\includegraphics[width=15cm]{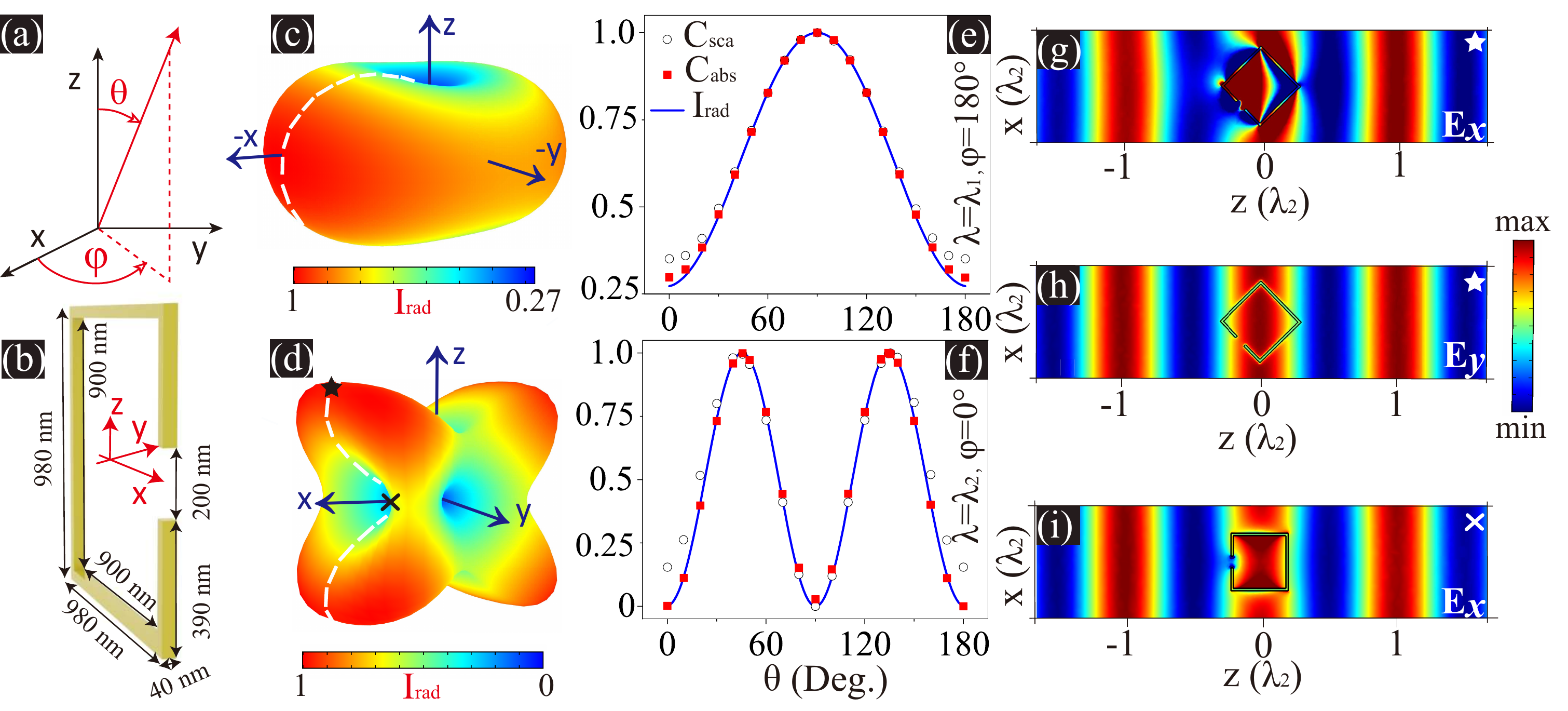}} \caption {\small (a) The spherical coordinate system with the azimuthal angle $\varphi$ and polar angle $\theta$. (b) A gold SRR with all geometric paramters and its orientation within the coordinate system specified. (c) and (d) Angular radiation patterns of the two QNMs (of eigenfrequencies $\tilde{\omega_1}$ and $\tilde{\omega_2}$) supported by the SRR, where the selected momentum-space semi-circles are marked by white-dashed lines. For the higher-order mode in (d), two points of maximum (star) and minimum (cross) radiation intensity have been marked. (e) and (f) Normalized radiation intensity or scattering (absorption) cross sections on the momentum-space semi-circles. (g) and (h) Near-field distributions (in terms of $\mathbf{E}_x$ and $\mathbf{E}_y$) for incident waves along the maximum-radiation direction marked in (d) with matched and orthogonal polarizations, respectively. (i) Near-field distributions (in terms of $\mathbf{E}_x$) for the incident wave along the minimum-radiation direction marked in (d) with the matched polarization.}\label{fig2}
\end{figure*} 

\section{Results and Discussions}
\begin{figure}[tp]
\centerline{\includegraphics[width=7.5cm]{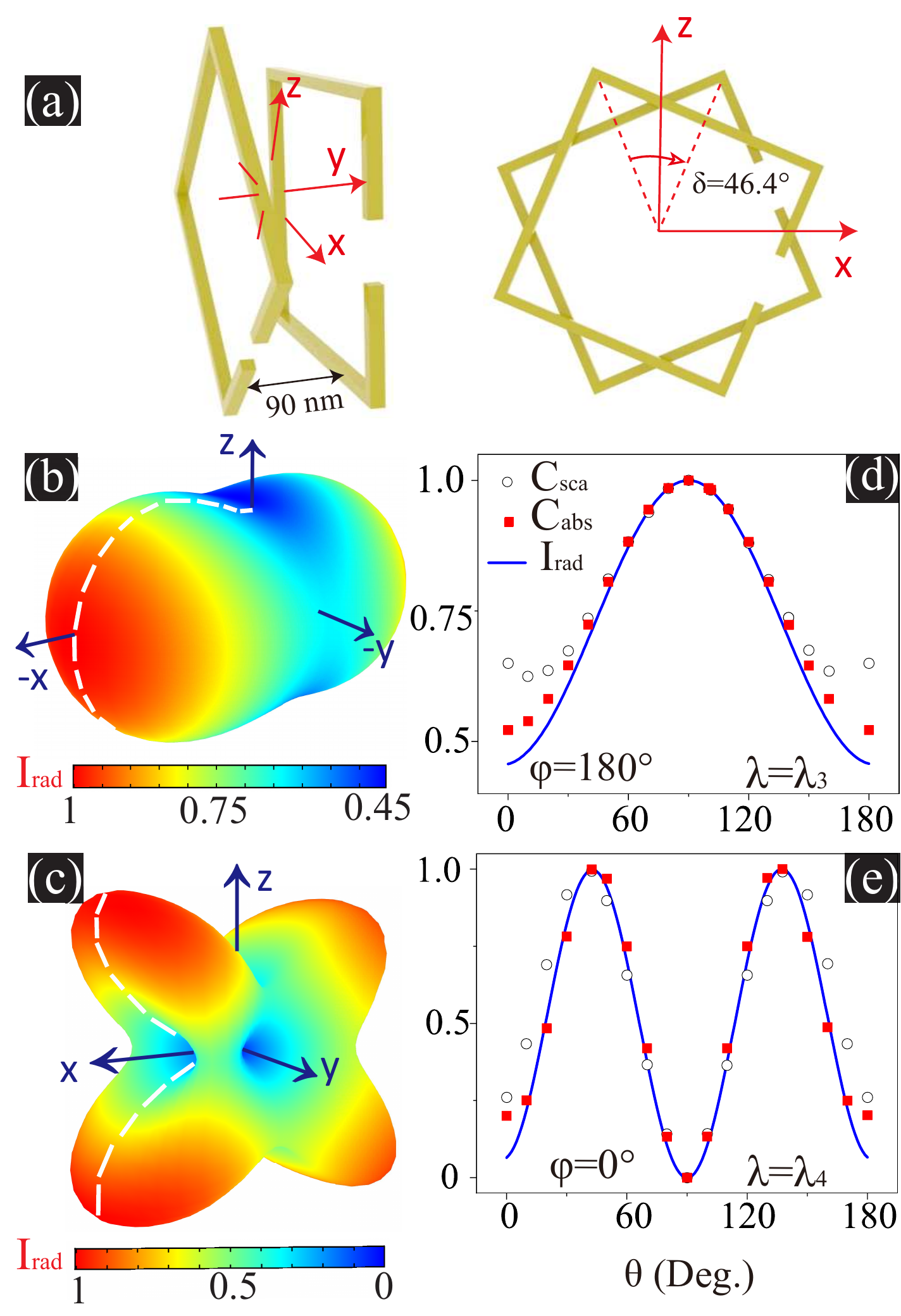}} \caption {\small (a) A pair of coupled parallel SRRs with a displacement of $90$ nm and a twist angle of $\delta=46.4^\circ$ along  the \textbf{y}-axis. Each SRR is identical to that shown in Fig.~\ref{fig2}(g). (b) and (c) Angular radiation patterns for the two selected QNMs (of eigenfrequencies $\tilde{\omega_3}$ and $\tilde{\omega_4}$) supported by the SRR pair, where the selected momentum-space semi-circles are marked by white-dashed lines. (d) and (e) Normalized radiation intensity or scattering (absorption) cross sections on the selected momentum-space semi-circles marked. }\label{fig3}
\end{figure}

To further verify the principles revealed, we employ the seminal structures of split ring resonators (SRRs), at the same time emphasizing that our theory is generally applicable to other photonic configurations. Throughout this work,  the SRRs investigated consist of gold with effective bulk permittivity fitted from data in Ref.~\cite{Johnson1972_PRB}, and numerical results are obtained using COMSOL Multiphysics. We start with an individual SRR, and all geometric parameters and its orientation with respect to the coordinate system shown in Fig.~\ref{fig2}(a) are specified in Fig.~\ref{fig2}(b). Two QNMs supported by this SRR are chosen with eigenfrequencies of $\tilde{\omega}_{1}=(2.326\times10^{14}-1.3258\times10^{13}\rm{i})$ rad/s (central resonant wavelength $\lambda_1=9.348~\mu$m) and $\tilde{\omega}_{2}=(8.055\times10^{14}-4.8841\times10^{12}\rm{i})$ rad/s ($\lambda_2=2.339~\mu$m), and their corresponding angular radiation patterns (in terms of ${\rm{I}}_{\rm{rad}}$) are shown in Figs.~\ref{fig2}(c) and (d), respectively. For each QNM, we have selected a semi-circle in the momentum-space [denoted by dashed white lines in Figs.~\ref{fig2}(c) \& (d)] which covers both directions of maximum and minimum radiation intensity throughout the momentum space. For calculations of scattering properties
 ($\rm C_{\rm{sca}}$ and $\rm C_{\rm{abs}}$) with incident waves along directions on those momentum-space semi-circles, we fix the angular frequency of the incident wave ${\omega}=\rm{Re}(\tilde{\omega})$ and make sure the incident polarization is matched to that of the radiation ($|\mathbf{J_{\rm{rad}}}\mathbf{J^{\dagger}_{\rm{inc}}}|=1$), reducing Eq.~(\ref{efficiency-dependence-2}) to:
\begin{equation}
\label{efficiency-dependence-3}
\rm C_{\rm{sca},\rm{ext},\rm{abs}} \propto {\rm{I}}_{\rm{rad}}.
\end{equation}
The normalized ${\rm{I}}_{\rm{rad}}$ and $\rm C_{\rm{sca,abs}}$ on our selected momentum-space semi-circles for both QNMs are shown respectively in Figs.~\ref{fig2}(e) and (f). Agreement between them is manifest confirming Eq.~(\ref{efficiency-dependence-3}),  despite some discrepancies along those directions of small ${\rm{I}}_{\rm{rad}}$. This is because small ${\rm{I}}_{\rm{rad}}$ is synonymous with small excitation efficiency [Eq.~(\ref{expansion-coefficient-jones})] and then contributions from other QNMs are not fully negligible any more, rendering our single-QNM approximation less accurate. The smaller is ${\rm{I}}_{\rm{rad}}$, the larger is the discrepancy, as is also the case for the results shown in  Figs.~\ref{fig3}(b) and (c) that will be discussed later.
We have also pinpointed the two directions of maximum and minimum radiation directions of the higher-order QNM, marked by a star ($\theta=44.3^\circ,~ \varphi=0$) and a cross ($\theta=90^\circ,~ \varphi=0$) in Fig.~\ref{fig2}(d), respectively. When the incident direction is along the maximum radiation direction, the near-field distributions for matched ($|\mathbf{J_{\rm{rad}}}\mathbf{J^{\dagger}_{\rm{inc}}}|=1$) and orthogonal ($|\mathbf{J_{\rm{rad}}}\mathbf{J^{\dagger}_{\rm{inc}}}|=0$) incident polarizations are shown respectively in Fig.~\ref{fig2}(g) (largest excitation efficiency and scattering)  and Fig.~\ref{fig2}(h) (close to zero excitation efficiency and scattering). Along the minimum radiation direction (${\rm{I}}_{\rm{rad}}\approx0$), it is shown in Fig.~\ref{fig2}(i) that, similar to that shown in Fig.~\ref{fig2}(h), scattering is almost absent despite the matched incident polarization. All those results are consistent with Eq.~(\ref{efficiency-dependence-2}), confirming that invisibility is obtainable for arbitrary incident directions with either ${\rm{I}}_{\rm{rad}}=0$ or $|\mathbf{J_{\rm{rad}}}\mathbf{J^{\dagger}_{\rm{inc}}}|=0$.

We then proceed to check the more sophisticated configuration of coupled SRRs schematically shown in Fig.~\ref{fig3}(a).  Here each SRR is identical to that shown in Fig.~\ref{fig2}(b) and they are parallel-displaced by $90$ nm and twisted by $\delta=46.4^\circ$ with respect to each other along the \textbf{y}-axis. Similarly, two QNMs supported by the SRR pair are selected with eigenfrequencies of $\tilde{\omega}_{3}=(1.7825\times10^{14}-1.1628\times10^{13}\rm{i})$ rad/s (central resonant wavelength $\lambda_3=10.568~\mu$m) and $\tilde{\omega}_{4}=(7.8753\times10^{14}-4.1295\times10^{12}\rm{i})$ rad/s ($\lambda_4=2.392~\mu$m), and their corresponding radiation patterns are demonstrated respectively in Figs.~\ref{fig3}(b) and (c). The normalized quantities (${\rm{I}}_{\rm{rad}}$ and $\rm C_{\rm{sca,abs}}$ with matched incident polarizations) on the selected momentum-space semi-circles [marked by white-dashed lines in Figs.~\ref{fig3}(b) and (c)] are shown in Figs.~\ref{fig3}(d) and (e), respectively. Similar to those  already observed in Figs.~\ref{fig2}(e) and (f), it is manifest in Figs.~\ref{fig3}(d) and (e) that: along directions with large ${\rm{I}}_{\rm{rad}}$ and thus also large coupling efficiencies, the agreement between both sets of results are excellent; while for small ${\rm{I}}_{\rm{rad}}$ and thus also low coupling efficiencies, obvious discrepancies arise, induced by non-negligible contributions from other QNMs.

\section{Conclusions and Perspectives}

We study momentum-space scattering extremizations with respect to varying incident directions from the perspective the QNMs.  When an individual QNM is effectively excited, we have revealed that: scattering is maximized for waves incident along the directions of maximum QNM radiations with matched radiation and incident polarizations; for an arbitrary direction of non-zero radiation, the scattering is fully eliminated with orthogonal radiation and incident polarizations; for incident directions along which there are no QNM radiations, the scatterings are null for arbitrary incident polarizations. We have effectively revealed the deep connections between QNM radiation patterns and momentum-space scattering evolutions, which are protected by the fundamental principles of  electromagnetic reciprocity and optical theorem, and thus can be naturally extended to other disciplines of wave physics.

In this study, we have confined our investigations to fully-polarized incident waves on the surface of the Poincar\'{e} sphere, and similar studies can be directly extended to the interior of the Poincar\'{e} sphere to encompass partially polarized and unpolarized incident waves. For example, for unpolarized incident waves, it is obvious that the scattering is maximized and minimized along directions of maximum and minimum QNM radiations, respectively.

We have shown in previous studies that the polarization distributions of the QNM radiations are globally bounded by the Poincar\'{e}-Hopf theorem~\cite{NEEDHAM__Visuala}, which secures the existence of polarization singularities and a global index sum of Euler characteristic $2$ of the momentum sphere~\cite{CHEN_Phys.Rev.Lett._Singularities,CHEN_2019_ArXiv190409910Math-PhPhysicsphysics_Linea}. In a similar manner, the maximum and minimum QNM radiation points (together with other critical points) are also globally bounded by the Morse theory~\cite{MATSUMOTO_2001__Introduction}, and explorations of those points and their Morse indexes from a photonics perspective can certainly further expand the richness of the vibrant field of topological physics.

The presence of discrepancies in Figs.~\ref{fig2}(e-f) and Figs.~\ref{fig3}(d-e) have already exposed the limitation of our theoretical model, which is only applicable to the single-QNM excitation regime. When there are symmetry-protected mode degeneracies or spectrally-close QNMs that are simultaneously co-excited, the principles revealed in this paper cannot be directly applied. A more encompassing model that can systematically solve the problem of the momentum-space scattering exterminations, for either multi-QNM scatterers or structured non-plane incident waves, is yet to be established.

\section*{acknowledgement}

This research was funded by the National Natural Science Foundation of China (12274462, 11674396, and 11874426), and the Science and Technology Planning Project of Hunan Province (2018JJ1033 and 2017RS3039).

\bibliographystyle{osajnl}
\bibliography{References_scattering3} 

\begin{thebibliography}{10}
\newcommand{\enquote}[1]{``#1''}

\bibitem{Bohren1983_book}
C.~F. Bohren and D.~R. Huffman, \emph{Absorption and Scattering of Light by
  Small Particles} (Wiley, 1983).

\bibitem{YARIV_2006__Photonics}
A.~Yariv and P.~Yeh, \emph{Photonics: {{Optical Electronics}} in {{Modern
  Communications}}} ({Oxford University Press}, {New York}, 2006), 6th ed.

\bibitem{LIU_2005__Photonic}
J.-M. Liu, \emph{Photonic {{Devices}}} ({Cambridge University Press},
  {Cambridge ; New York}, 2005).

\bibitem{ZUMOFEN_Phys.Rev.Lett._perfect_2008-1}
G.~Zumofen, N.~M. Mojarad, V.~Sandoghdar, and M.~Agio, \enquote{Perfect
  {{Reflection}} of {{Light}} by an {{Oscillating Dipole}},} Phys. Rev. Lett.
  \textbf{101}, 180404 (2008).

\bibitem{GOUESBET_generalized_2011}
G.~Gouesbet and G.~Gr{\'e}han, \emph{Generalized {{Lorenz}}-{{Mie Theories}}}
  ({Springer Science \& Business Media}, 2011).

\bibitem{SHEN_2019_LightSciAppl_Optical}
Y.~Shen, X.~Wang, Z.~Xie, C.~Min, X.~Fu, Q.~Liu, M.~Gong, and X.~Yuan,
  \enquote{Optical vortices 30 years on: {{OAM}} manipulation from topological
  charge to multiple singularities,} Light Sci Appl \textbf{8}, 1--29 (2019).

\bibitem{WAN_2022_Natl.Sci.Rev._Photonic}
C.~Wan, J.~Chen, A.~Chong, and Q.~Zhan, \enquote{Photonic orbital angular
  momentum with controllable orientation,} Natl. Sci. Rev. \textbf{9}, nwab149
  (2022).

\bibitem{FORBES_Nat.Photonics_Structured}
A.~Forbes, M.~{de Oliveira}, and M.~R. Dennis, \enquote{Structured light,} Nat.
  Photonics \textbf{15}, 253--262 (2021).

\bibitem{RUDIN_1976__Principles}
R.~W. Rudin, \emph{Principles of {{Mathematical Analysis}}} ({McGraw-Hill
  Publishing Company}, {Auckland}, 1976), 3rd ed.

\bibitem{LALANNE__LaserPhotonicsRev._Light}
P.~Lalanne, W.~Yan, K.~Vynck, C.~Sauvan, and J.-P. Hugonin, \enquote{Light
  {{Interaction}} with {{Photonic}} and {{Plasmonic Resonances}},} Laser
  Photonics Rev. \textbf{12}, 1700113 (2018).

\bibitem{CHEN_Phys.Rev.Lett._Extremize}
W.~Chen, Q.~Yang, Y.~Chen, and W.~Liu, \enquote{Extremize {{Optical
  Chiralities}} through {{Polarization Singularities}},} Phys. Rev. Lett.
  \textbf{126}, 253901 (2021).

\bibitem{DEHOOP_Appl.sci.Res._reciprocitya}
A.~T. {de Hoop}, \enquote{A reciprocity theorem for the electromagnetic field
  scattered by an obstacle,} Appl. Sci. Res. \textbf{8}, 135--140 (1960).

\bibitem{CHEN_2021_Phys.Rev.B_Arbitrary}
W.~Chen, Q.~Yang, Y.~Chen, and W.~Liu, \enquote{Arbitrary
  polarization-independent backscattering or reflection by rotationally
  symmetric reciprocal structures,} Phys. Rev. B \textbf{103}, 045422 (2021).

\bibitem{SOUNAS_Opt.Lett.OL_extinction_2014}
D.~L. Sounas and A.~Al{\`u}, \enquote{Extinction symmetry for reciprocal
  objects and its implications on cloaking and scattering manipulation,} Opt.
  Lett. \textbf{39}, 4053--4056 (2014).

\bibitem{CHEN_2020_Phys.Rev.Research_Scatteringa}
W.~Chen, Q.~Yang, Y.~Chen, and W.~Liu, \enquote{Scattering activities bounded
  by reciprocity and parity conservation,} Phys. Rev. Research \textbf{2},
  013277 (2020).

\bibitem{CHEN_Phys.Rev.Lett._Singularities}
W.~Chen, Y.~Chen, and W.~Liu, \enquote{Singularities and {{Poincare}}
  {{Indices}} of {{Electromagnetic Multipoles}},} Phys. Rev. Lett.
  \textbf{122}, 153907 (2019).

\bibitem{Johnson1972_PRB}
P.~B. Johnson and R.~W. Christy, \enquote{Optical constants of the noble
  metals,} Phys. Rev. B \textbf{6}, 4370 (1972).

\bibitem{NEEDHAM__Visuala}
T.~Needham, \emph{Visual {{Differential Geometry}} and {{Forms}}: {{A
  Mathematical Drama}} in {{Five Acts}}} ({Princeton University Press},
  {Princeton}, 2021).

\bibitem{CHEN_2019_ArXiv190409910Math-PhPhysicsphysics_Linea}
W.~Chen, Y.~Chen, and W.~Liu, \enquote{Line {{Singularities}} and {{Hopf
  Indices}} of {{Electromagnetic Multipoles}},} Laser Photonics Rev.
  \textbf{14}, 2000049 (2020).

\bibitem{MATSUMOTO_2001__Introduction}
Y.~Matsumoto, \emph{An {{Introduction}} to {{Morse Theory}}} ({Amer
  Mathematical Society}).

\end{thebibliography}

\end{document}